\documentclass[%
reprint,
superscriptaddress,
 amsmath,amssymb,
 aps,
 prx,
floatfix,
]{revtex4-2}

\usepackage[english]{babel}

\usepackage[hidelinks]{hyperref}
\usepackage{graphicx}
\usepackage{dcolumn}
\usepackage{bm}
\usepackage{comment}
\usepackage{physics}
\usepackage{cleveref}
\usepackage{xfrac}

\newcommand{\probgiven}[2]{\ensuremath{\operatorname{Pr}\left(#1 \middle| #2\right)}}
\newcommand{\probguess}[1]{\ensuremath{\operatorname{Pr}_g\left(#1\right)}}

\newcommand{\linbo}{{$ \text{LiNbO}_3 $} }

\begin{document}

\preprint{APS/123-QED}

\title{High-dimensional detection-loophole-free measurement-device-independent quantum random number generator}

\author{Joakim Argillander}
\email{joakim.argillander@liu.se}
\thanks{These authors contributed equally.}
\author{Daniel Spegel-Lexne}
\thanks{These authors contributed equally.}

\author{Martin Clason}%

\affiliation{%
 Department of Electrical Engineering\\
 Linköping University, SE-581 83, Linköping, Sweden
}%

\author{Pedro R. Dieguez}

\author{Marcin Paw\l{}owski}
\affiliation{
International Centre for Theory of Quantum Technologies \\
University of Gda\'{n}sk, Jana Bazynskiego 8, 80-309 Gda\'{n}sk, Poland
}%

\author{Anubhav Chaturvedi}
\email{anubhav.chaturvedi@pg.edu.pl}
\affiliation{
International Centre for Theory of Quantum Technologies \\
University of Gda\'{n}sk, Jana Bazynskiego 8, 80-309 Gda\'{n}sk, Poland
}%
\affiliation{
Faculty of Applied Physics and Mathematics \\
Gda\'{n}sk University of Technology, Jana Bazynskiego 8, 80-309 Gda\'{n}sk, Poland
}%

\author{Guilherme B. Xavier}
\email{guilherme.b.xavier@liu.se}
\affiliation{%
 Department of Electrical Engineering\\
 Linköping University, SE-581 83, Linköping, Sweden
}%

\date{\today}

\begin{abstract}
Certifying random number generators is challenging, especially in security-critical fields like cryptography. Here, we demonstrate a measurement-device-independent quantum random number generator (MDI-QRNG) using high-dimensional photonic path states. Our setup extends the standard qubit beam-splitter QRNG to a three-output version with tunable fiber-optic interferometers acting as tunable beam splitters and superconducting detectors. This setup generates over 1.2 bits per round and 1.77 Mbits per second of certifiably secure private randomness without requiring \emph{any} trust in the measurement apparatus, a critical requirement for the security of real-world cryptographic applications. Our results demonstrate certifiably secure high-dimensional quantum random-number generation, paving the way for practical, scalable QRNGs without the need for complex devices.

\end{abstract}

\maketitle


\section{Introduction}
Randomness is a vital resource in many fields of physics, engineering, and computer science. Particularly, in cryptography, random numbers are used to generate keys for encryption and decryption, and as random tokens \cite{ma_quantum_2016, herrero-collantes_quantum_2017}. The strength of any cryptographic system is directly dependent on the difficulty of finding the encryption key. If a key has been generated using a predictable process, an adversary can more easily find the key and break the encryption. Therefore, it is crucial that the random numbers used in cryptography are truly random and private. In this case, it is not only sufficient that the generated bits are random, but also that they are unpredictable by an adversary \cite{acin_certified_2016}. Therefore, it is important to highlight a clear difference between the notion of randomness and privacy, where the former does not necessarily imply the latter. Private randomness is a stronger notion than traditional randomness as it also safeguards against an adversary that has access to the random number generator (RNG) and can influence the output. Adversaries can be either malicious or unintentional, where the latter can be caused by hardware faults or environmental noise. A particular critical case is where the malicious adversary is the manufacturer of an RNG themselves, as they have full control over the inner workings of the RNG device. In principle, an antagonistic manufacturer could have pre-programmed into the RNG device a sequence of algorithmically generated numbers. This completely eliminates the need for the manufacturer to attempt to predict the output of the RNG, as it is already fully known to them. This work does not distinguish between the two cases of adversaries, and we will use the term \emph{private randomness}\cite{acin_certified_2016} to refer to the property of a random number generator that guarantees that the output is unpredictable by a modeled adversary, Eve. 

Quantum mechanics provides a way to generate private randomness by exploiting the inherent unpredictability of quantum systems. Quantum random number generators (QRNGs) use quantum phenomena, such as the measurement of quantum states, to generate random numbers that can be made unpredictable and private \cite{ma_quantum_2016, herrero-collantes_quantum_2017}. The first developed QRNGs focused on performing measurements on individual quantum systems, of which the bits are assigned to the random outcomes of these measurements \cite{Jennewein2000, Stefanov2000, Dynes2008, Wayne2009, Nie2014}. In this case, private randomness is only possible with complete trust in the physical devices comprising the QRNG, called the device-dependent (DD) approach. Advantages of DD-QRNG compared to other random number generators are limited to only allowing higher bitrates at the cost of weaker privacy guarantees. However, it is possible to build QRNGs with no knowledge or trust in the inner workings of the physical devices, which is based on the device-independent (DI) framework \cite{Pironio2010, Bierhorst2018, Liu2018, Zhang2020, Shalm2021}. 

The DI approach has very high experimental requirements, mainly stemming from the necessity of a (detection) loophole-free Bell inequality violation \cite{Hensen_2015, Giustina_2015, Shalm_2015, Rosenfeld_2017,PhysRevLett.129.230403,Chaturvedi2024,Gigena2025}, which makes the DI-QRNGs impractical for most real-world applications. The experimental complexity also leads to comparatively low randomness generation rates, which is a significant drawback for practical implementations. Interestingly, it is possible to obtain partial device-independence with the relaxation to a restricted set of natural assumptions which constitute a semi-device-independent (SDI) framework. It was first shown that DI security is possible given trust in an upper-bound on the dimension of the transmitted quantum state \cite{pawlowski2011}. The assumptions needed for SDI-QRNGs are usually much more reasonable than device-dependent QRNGs, thus, great interest has appeared in the last years in the experimental implementation of SDI-QRNGs \cite{Lunghi2015, Cao2016, nie_experimental_2016, Brask2017, Li2019, Rusca2019, carine_multi-core_2020, Drahi2020, Mironowicz2021, Pivoluska2021, Lin2022, Liu2023,argillander_all-fiber_2022, alarcon_dynamic_2023, Zhao2024, Bertapelle2025, argillander_qutrit_cleo_2025}. SDI-QRNGs that do not place assumption on the measurement devices are called measurement-device-independent (MDI) QRNGs \cite{chaturvedi_measurement-deviceindependent_2015}. MDI-QRNGs only require that the source of quantum states is well characterized, and thus, no assumptions are needed on the measurement apparatus (including the noisy and lossy channel). 
Beyond mitigating detector side channels \cite{Xu2020}, the MDI paradigm is a natural stepping stone to DI security in light of the recently proposed routed Bell experiments \cite{Chaturvedi2024}. In such tests, a switch randomly routes some rounds to a short path where high-efficiency devices witness a strong Bell violation; these rounds self-test the source (and nearby measurements). The same source is then used in long-path rounds for randomness generation with uncharacterized, lossy measurement devices, so that the usual MDI assumption on preparations may eventually be replaced by DI certification of the effective preparations from the short-path, while leaving the lossy measurement hardware unaltered.

In MDI-QRNGs the certification of private randomness is provided by the measurement results of so-called test states that the source produces in order to probe the measurement device. Previous MDI-QRNGs have employed telecom wavelength time-bin quantum states \cite{nie_experimental_2016}, transverse spatial in two-dimensional states using few-mode fibers \cite{alarcon_dynamic_2023}, and also expanded the encoding to higher-dimensions with multi-core fibers \cite{carine_multi-core_2020}. Another setup was based on performing unambiguous state discrimination \cite{Brask2017}, while a more recent result demonstrated the use of polarization quantum states at near infra-red wavelengths produced from a perovskite light emitting diode \cite{Argillander2023}.

In this paper, we improve on the randomness certification of previous MDI-QRNGs by certifying randomness without being affected by the detection loophole. This is achieved through the use of a conceptually simple and dynamic state preparation scheme connected to untrusted measurement operations combined with high-efficiency superconducting nanowire single-photon detectors (SNSPDs). Our QRNG employs path-encoded quantum states produced from heavily attenuated laser pulses, which form a source of weak coherent states (WCS). To further boost the randomness certification per measurement round, we implement our QRNG with three-dimensional path states, and through an optimization of the average photon number per pulse, we show a certified randomness generation rate of more than $1.2$ bits/round, clearly beating the qubit limit of 1 bit/round. An important benefit of the protocol presented in this work is that it can be used for real-time certification of the QRNG during operation. Online certification allows for continuous monitoring of the QRNG before the generated random numbers are used in a cryptographic protocol, which is an important feature for integration into realistic systems. Our setup is furthermore directly scalable to even higher dimensions, opening up the possibilities for further boosting the randomness certification rate. Our results show an effective way to generate a high-efficiency certified randomness generation rate with a practical and scalable setup that can have many applications within quantum communication and information technologies. Since, the trust in the preparations in the MDI setting can replaced by short path device independent self-testing via routed Bell tests \cite{Chaturvedi2024,Lobo2024certifyinglongrange,chaturvedi2025extendingquantumcorrelationsarbitrary}, our results demonstrate the utility of higher-dimensional quantum systems in practical device independent devices for quantum security.

\section{Results}
\label{sec:results}
The branching-path QRNG is arguably one of the most well-known generators, and is conceptually very simple \cite{herrero-collantes_quantum_2017}. The output of a source of single-photon states is divided into two outputs, usually with a beamsplitter (BS) with its two outputs connected to single-photon detectors \cite{Jennewein2000, Stefanov2000}. A single-photon impinging on the beamsplitter has a $50/50$ probability of being routed to either of its outputs. Quantum mechanically, the output state of the beamsplitter is a superposition of the two outputs, which can be written as
\begin{equation}
	\ket{\psi} = \frac{1}{\sqrt{2}} \left(\ket{\mathrm{0}} + \ket{\mathrm{1}} \right),
\end{equation}
where $\ket{\mathrm{0}}$ and $\ket{\mathrm{1}}$ are the two output modes of the beamsplitter. Whenever a detection occurs in either output, the corresponding logical bit ('$0$' or '$1$') is recorded by the electronics connected to the single-photon detectors. In the event that both detectors register detections, which can occur from either multi-photon emission from the source or from background or dark counts on the detectors, the round is typically discarded. Likewise, rounds where no photon detection was registered, i.e., a no-click, are typically discarded.

\begin{figure*}[ht!]
	\centering
	\includegraphics[trim={0.9cm 22.5cm 0.7cm 1cm},clip,width=\textwidth]{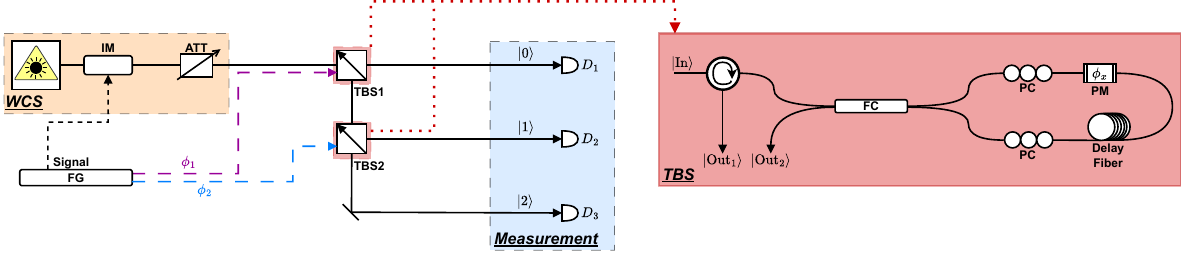}
	\caption{Experimental setup. A source of weak coherent states (WCS) prepares path-encoded qutrit states using two tunable beam splitters (TBSs). A function generator (FG) controls the intensity modulator (IM) used to chop the continuous-wave (CW) laser into 10 ns wide pulses, which are then attenuated to the single-photon level using a series of variable optical attenuators (ATT). The two tunable beamsplitters (TBSs) are each actually comprised of a fiber-optical Sagnac interferometer (inset), containing a phase modulator (PM) $\phi_x$, an optical fiber delay and manual polarization controllers (PC). The internal relative phase $\phi_x$ in each interferometer TBS controls the splitting ratio of the TBS. The three outputs of the TBSs are connected to three superconducting nanowire single-photon detectors (SNSPDs).}
\label{fig:qutrit-expsetup}
\end{figure*}

We take the branching-path QRNG as the basis for our experiment, and upgrade to three dimensions by cascading two beamsplitters together such that now three outputs are possible from the same input (\Cref{fig:qutrit-expsetup}). In order to remove the explicit trust in the measurement apparatus, we employ the MDI-protocol \cite{nie_experimental_2016, supic_measurement-device-independent_2017} where the source needs to dynamically prepare a set of different states that are used for testing the measurement device, and a single state for generating the random numbers. The test states are used to certify the amount of private randomness generated by the QRNG, while the generation state is used to generate randomness. When the device is preparing and measuring either of the test states, the device is said to operate in test mode (T), while the generation state is prepared and measured in generation mode (G). 

The T mode states consist of the set of eigenstates $\{\ket{\psi_T}\}=\ket{\psi_1}, \ket{\psi_2},\ldots, \ket{\psi_d} $ of the measurement operator $\mathcal{M}_d$, while the G mode employs one state that corresponds to a linear superposition of all the T eigenstates $\ket{\psi_G} = \frac{1}{\sqrt{d}} \sum \alpha_i e^{\phi_i}\ket{\psi_i}$, where $\alpha_i$ are the amplitude coefficients and $\phi_i$ the relative phase between the state components. Measurement outcomes for each round are recorded separately for each mode, with the T mode detections  used to certify the generated randomness within the G mode. During operation, a user can switch between the T and G modes, which allows for real-time certification of the QRNG. In order to simulate a user's involvement in the operation of the QRNG, we use a pseudo-random number generator (PRNG) to make the choice. The PRNG is biased such that the T mode is used $3\%$ of the time, while the G mode is used $97\%$ of the time. Since the T mode does not generate randomness, using it too often reduces the overall rate of random number generation. Conversely, too few test rounds introduce statistical uncertainty, which can prevent the security analysis from determining the certifiable randomness.  

In our experiment we switch between the T and G modes, and fine-tune the $\alpha_i$ coefficients with a fully fiber-based setup (\Cref{fig:qutrit-expsetup} inset) consisting of two tunable beamsplitters (TBSs). The TBSs themselves are implemented with fiber-optical Sagnac interferometers \cite{alarcon_dynamic_2023, argillander_tunable_2022} (see \Cref{appendix:methods-sagnac} for more details). We employ a continuous-wave (CW) telecom fiber laser (NKT Photonics X15) with a center wavelength of $\lambda = 1550.12\text{ nm}$ which we chop up into pulses $10 \text{ ns}$ wide using a $10\text{ GHz}$ telecom lithium niobate ($\textrm{LiNbO}_3$) intensity modulator at a repetition rate of $2.2 \text{ MHz}$, produced from a home-made Field Programmable Array (FPGA)-based pulse generator. The FPGA synchronizes the operation of the QRNG with the detectors. The pulses are then attenuated to single-photon level using two cascaded electrically controlled variable optical attenuators (ATTs) in order to create a source of weak coherent pulses (WCPs) before the measurement operation with average photon number $\mu$ per pulse, where the probability of a pulse containing $n$ photons is $(\mu^n e^{-\mu})/n!$ \cite{fox_quantum_2006}. The WCPs are then sent to the experimental setup, where they are split into three paths using the two cascaded TBSs. The first TBS is controlled by \linbo phase modulator $\phi_1$, which allows us to dynamically tune the splitting ratio of the TBS. The splitting ratio of the second TBS is controlled by the second \linbo phase modulator $\phi_2$. The outputs of the two TBSs are connected to three single-photon detectors, which are used to measure the output state of the QRNG.

\begin{figure}
    \centering
    \includegraphics[width=\linewidth]{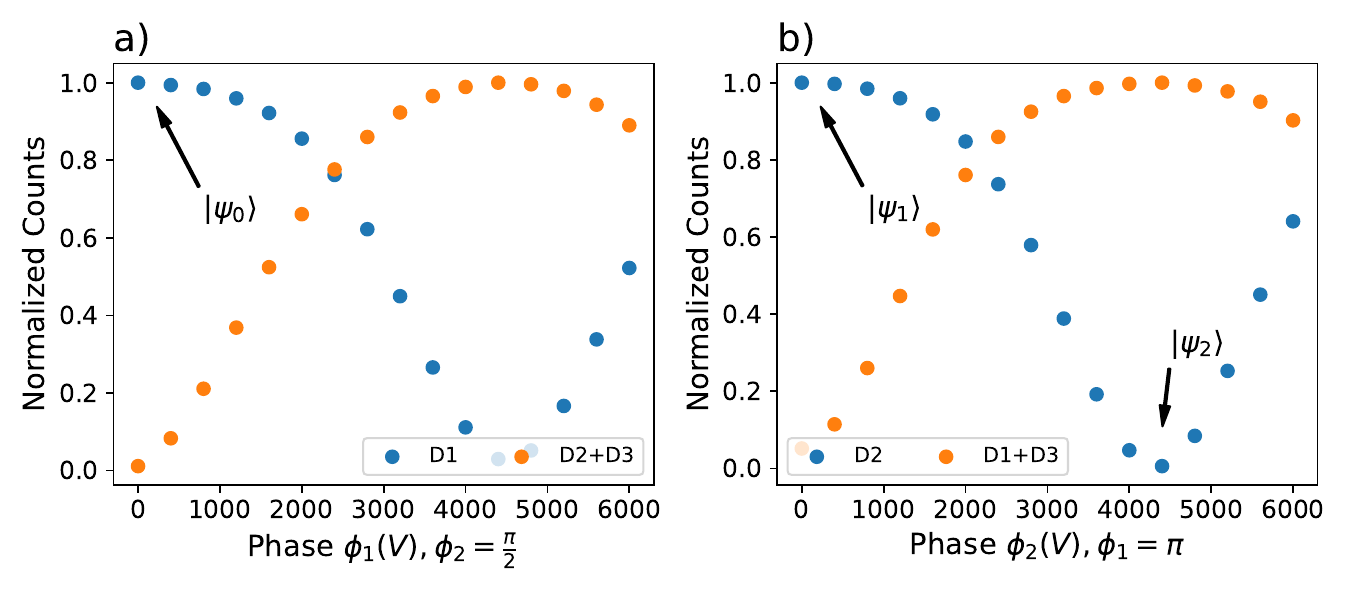}
    \caption{Interference patterns of the first a) and second b) Sagnac interferometer as a function of the voltages applied to the phase modulators $\phi_1$ and $\phi_2$. Also highlighted are the settings that correspond to the test states $\ket{\psi_0}, \ket{\psi_1}$ and $\ket{\psi_2}$}
    \label{fig:qutrit-interference-patterns}
\end{figure}

We characterize the cascaded interferometer setup by measuring the interference patterns as a function of the two phases $\phi_1$ and $\phi_2$ (\Cref{fig:qutrit-interference-patterns}). The patterns are measured by scanning the two phases (scanning the voltages applied to $\phi_1$ and $\phi_2$) and recording the counts at the three detectors. The interference patterns show that the output probabilities of the three detectors can be tuned to be equal, which corresponds to the prepared state $\ket{\psi_G}$, which is the state used for the G mode. The interference patterns also show that the prepared test states $\left\{\ket{\psi_T}\right\}$ can be tuned to be orthogonal to each other, which is a requirement for the MDI-QRNG protocol. 

Following the protocol in in \cite{supic_measurement-device-independent_2017} we alternate between preparing and measuring states for randomness generation $(\ket{\psi_G})$ and the eigenstates ($\ket{\psi_x}\in\{\ket{0}, \ket{1}, \ket{2}\}$) of the measurement operator $\mathcal{M}$ used for the T rounds. The experiment is divided into blocks of approximately $2.2 \times 10^6$ rounds, each lasting $1$ s. For each block, the source prepares the same state randomly chosen from the three test states or the randomness generation. We use a PRNG to select which state with a bias of $p(\ket{\psi_G}) = 0.97$, $p(\ket{\psi_0}) = p(\ket{\psi_1}) = p(\ket{\psi_2})= 0.01 $ to prepare and measure either state. For the test rounds each prepared state yields an outcome $a$. We then experimentally estimate the probabilities 


The MDI protocol is then run continuously during a period of approximately two hours, and we measure a stable certified generation rate of $1.77 \text{ Mbps}$ (\Cref{fig:qutrit-results}a), employing an optimized average mean photon number $\mu = 1.22$ (see \Cref{appendix:methods-certification}).  We also plot in \Cref{fig:qutrit-results}b the detection probability for the three test states during the same run as $\sfrac{\psi_i}{\sum\psi_i}$. The test state probabilities are not uniform in respect to the runtime of the experiment, as they are randomly chosen during the run, following the MDI-QRNG protocol. We obtain an average success probability of $(0.999\pm1.3\cdot10^{-7}, 0.986\pm2\cdot10^{-6}, 0.990\pm2\cdot10^{-6})$ for the three test states $(\ket{0}, \ket{1}, \ket{2})$, respectively, and they remain stable throughout the experiment. The asymmetry in detection probabilities stems from the fact that the imperfect splitting ratio in the first beamsplitter limits the performance of the second.

\begin{figure}[ht]
\centering
\includegraphics[width=\linewidth]{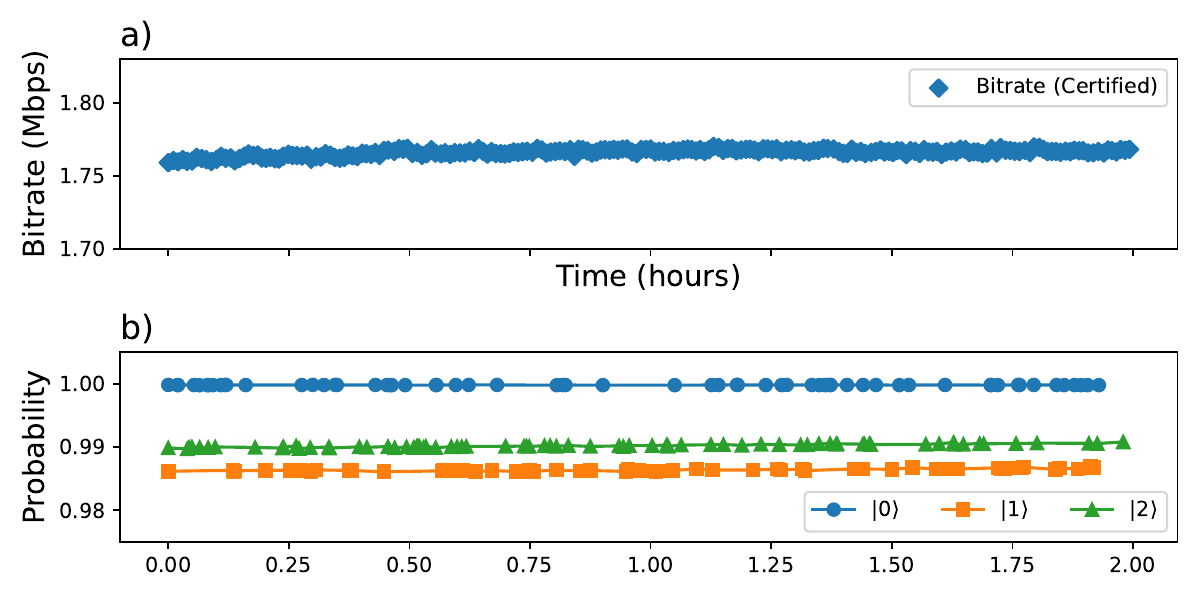}
\caption{Experimental results of the qutrit QRNG. a) Certified bitrate over time. The certified bitrate is given by the detection rate (i.e. the symbol rate), multiplied with the amount of certified randomness per round. b) Success probability over time, of the test states, needed to provide randomness certification using the MDI protocol. Each point corresponds to one test state block, and they are not uniformly distributed since they are randomly chosen throughout the experiment (please see \Cref{sec:results} and \Cref{appendix:methods-experimental-qutrit-mdi-qrng}).}
\label{fig:qutrit-results}
\end{figure}

We employed high-efficiency SNSPDs (idQuantique ID281) in order to maximize the randomness certified rate in light of the detection loophole (see \Cref{appendix:methods-detection-loophole}). The nominal system efficiencies are $93.2\%$, $92\%$, and $80\%$ for D$_0$, D$_1$ and D$_2$ respectively. The mean dark count rate is $19$, $9$, and $1.3$ counts per second for the three detectors. As the detectors are directly connected to the outputs of the measurement operator in the path-basis, the only additional loss comes from fiber connectors to patch cords leading to the detector cryostat, which is installed in an adjacent room to the lab where the experiment is performed. Taking these losses into account, the effective efficiencies are $86.2\%$, $90.0\%$, $75.1\%$ respectively, which corresponds to the total detection efficiency from the source device to the detector (including the measurement apparatus).

\begin{figure}[htbp]
\centering
\includegraphics[width=\linewidth]{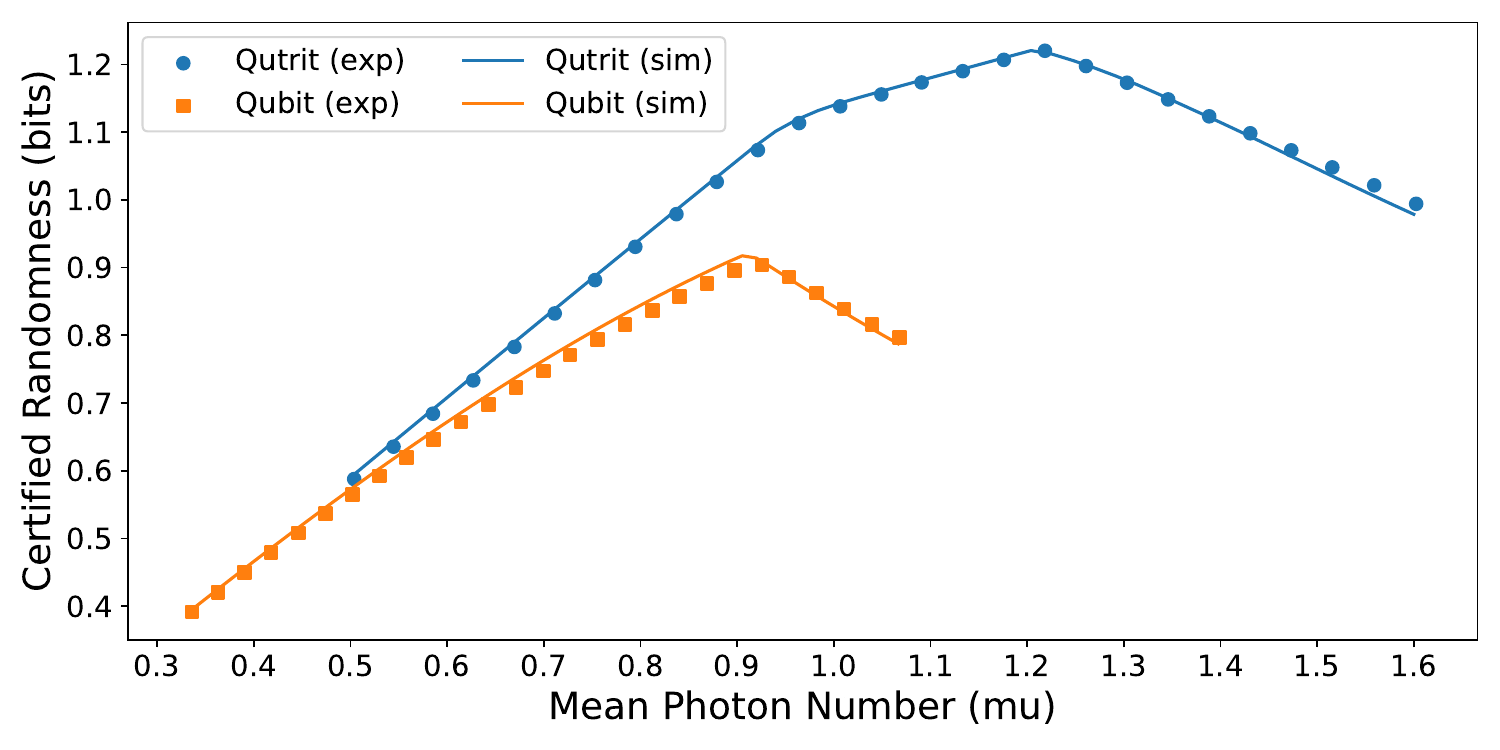}
\caption{Certified randomness per experimental round (data points) for the qutrit and qubit case, along with simulated certified randomness (line) for different mean photon numbers (see \Cref{appendix:methods-simulations}). The experimental qubit (qutrit) cases achieves a maximum $0.92$ ($1.22$) certified bits/round at $\mu=0.91$ ($\mu=1.22$), taking into account the measured test state probabilities. This clearly demonstrates the advantage of employing higher-dimensional quantum states for certified randomness generation.\label{fig:qutrit-certified-randomness}}
\end{figure}

We maximize the randomness generation by including in the certification procedure not only the single detection events at the three outputs, but also the double and triple click events, which stem from multi-photon emission coming from the photon number distribution of the WCSs. Dark count events also contribute to the randomness generation, but these are negligible given the detection rate of the experiment. To analyse the dependence on the randomness certification rate on the average photon number $\mu$ per pulse, we perform a post-processing step on the recorded time-tagged data of the experimental run. This step consists of applying different detection windows on the time-tagged data (using an Id Quantique ID1000 time tagger with 1 ps timing resolution), corresponding to different values of $\mu$. The detection statistics are then used to calculate the certified randomness per experimental round using the method described in \cite{supic_measurement-device-independent_2017} (see \Cref{appendix:methods-certification} and \Cref{appendix:methods-finite-round-correction}). Intuitively, a low $\mu$ lowers the overall detection rate as there is a higher probability that the vacuum state is sent, while a too high value for $\mu$ causes the detectors to saturate, and thus no randomness is produced. Using the randomness certification method described in \Cref{appendix:methods-certification}, we employ the detection statistics for each post-processed $\mu$ and calculate the certified randomness per experimental round, plotted in \Cref{fig:qutrit-certified-randomness}.

In order to clearly demonstrate the quantum advantage of employing high-dimensional states, we select from the measured data, through binning of the measurement outcomes, only the outputs of detectors D$_1$ and D$_2$. This effectively forms a standard branching-path QRNG of dimension $d=2$ (see \Cref{appendix:methods-binning}). We then perform the same procedure as for the qutrit case and calculate the generated certified randomness (\Cref{fig:qutrit-certified-randomness}). We observe that in the case of the binned qubit QRNG, the certified randomness is lower than the qutrit case, and the maximum is reached at $\mu = 0.91$. The maximum certified randomness for the binned qubit QRNG is $0.92$ bits/round, which is significantly lower than the $1.2$ bits/round obtained with the qutrit QRNG. This clearly demonstrates the advantage of employing higher-dimensional states in QRNGs, as they allow for a higher certified randomness generation rate. We note that the maximum information content in a qutrit is $\log_2(3) \approx 1.585$ bits, compared to the qubit case where the information content is $\log_2(2) = 1$ bit, with the differences between the theoretical and experimental maximums coming from the limited visibility in the test state sucess probabilities. 
\section{Discussion}
Although device-dependent quantum random number generators are very mature with many commercial products available, the issue of randomness certification and privacy is far too dependent on the manufacturer. The device-independent approach provides the strongest possible certification, but it remains completely impractical for applications for the foreseeable future. Semi-device-independent QRNGs are able to provide a higher degree of certification than standard QRNGs while having only modestly higher experimental requirements. Therefore, SDI-QRNGs hold great promise to replace standard QRNGs for practical applications. 

We have implemented a high-dimensional measurement-device-independent QRNG, which was able to clearly beat the benchmark value of $1$ bit of certified randomness per round. Furthermore, when combined with high-efficiency single-photon detectors together with an efficient measurement scheme, our setup also closed the detection loophole, thus providing stronger randomness certification. Our scheme is based on a fully fiber-optical platform, which provides a stable setup which also allowed very high detection probability for the test states, a crucial requirement for the high value of certified randomness obtained. For the dynamic state preparation for the MDI protocol, we implemented tunable beamsplitters with Sagnac interferometers. Our results show a way forward for building practical QRNGs while providing a high rate of certified randomness that only depend on limited assumptions. We have shown that the certified randomness generation rate can be further improved by employing higher-dimensional quantum states, such as qutrits, which is a promising direction for future work. In comparison to methods based on unambiguous state discrimination in \cite{Brask2017}, we are able to achieve a higher certified randomness generation rate, even in the case of a qubit QRNG. This is due to the fact that we are able to use the full photon number statistics of the WCS, which allows us to certify randomness from multi-photon events. 

\section*{Appendix}
\appendix
\label{appendix:methods}
\section{Fiber-optical tunable beam splitters}
\label{appendix:methods-sagnac}
The tunable beamsplitters (TBSs) are implemented with fiber-optical Sagnac interferometers, which are based on the interference of two counter-propagating pulses in a fiber loop. The Sagnac loop is formed by a fiber-optical coupler, which splits the incoming pulse into two counter-propagating pulses. The two pulses then travel in opposite directions around the loop and recombine at the same coupler. The relative phase between the two pulses can be controlled by introducing a phase shift $\phi$ using a phase modulator in one of the arms of the loop, before the two pulses recombine at the coupler. In order not to subject both counter-propagating wave packets to the same phase modulation (which would result in a 0 net phase shift), we add an asymmetric delay line ($\lessapprox5$ meter) in the Sagnac loop, which effectively ensures that phase modulation is only added to the clockwise propagating wave packet. The phase modulator is driven by a voltage signal, which can be adjusted to control the relative phase shift between the two counter-propagating pulses. An incoming pulse creates the superposition state $\ket{\psi} = \alpha\ket{\text{CW}} + (1-\alpha) e^{i\phi}\ket{\text{CCW}}$, where $\ket{\text{CW}}$ and $\ket{\text{CCW}}$ are the clockwise and counter-clockwise modes of the Sagnac loop, respectively. The output probabilities for the $\ket{\text{CW}}$ and $\ket{\text{CCW}}$ directions are proportional to $\mathrm{cos}^2(\frac{\phi}{2})$ and $\mathrm{sin}^2(\frac{\phi}{2})$ respectively. This allows for continuous adjustment of the output probabilities between the two ports by adjusting the relative phase $\phi$ between the counter-propagating pulses, effectively forming a tunable beamsplitter \cite{argillander_tunable_2022}.

The first TBS is controlled by phase $\phi_1$, and while the second TBS is implemented in the same way, it is governed by a different phase modulator $\phi_2$, which is independently controlled. The two TBSs are cascaded such that the output probabilities of the three outputs are proportional to $\mathrm{cos}^2(\frac{\phi_1}{2})\mathrm{cos}^2(\frac{\phi_2}{2})$, $\mathrm{sin}^2(\frac{\phi_1}{2})\mathrm{cos}^2(\frac{\phi_2}{2})$ and $\mathrm{sin}^2(\frac{\phi_1}{2})\mathrm{sin}^2(\frac{\phi_2}{2})$ respectively. At the three outputs of the two cascaded TBSs, the following three-dimensional path state is produced 
\begin{equation}
	\label{eq:qutrit-full-state}
	\ket{\psi} = \alpha\ket{0} + e^{\varphi_1}\beta\ket{1} + \gamma e^{\varphi_2}\ket{2},
\end{equation}
where 
\begin{equation}
	\begin{split}
		\alpha &= \mathrm{cos^2(\frac{\phi_1}{2})}, \\
		\beta &= \mathrm{sin^2}(\frac{\phi_1}{2})\mathrm{cos^2}(\frac{\phi_2}{2}), \\
		\gamma &= \mathrm{sin^2}(\frac{\phi_1}{2})\mathrm{sin^2}(\frac{\phi_2}{2}).
	\end{split}
\end{equation}

As shown in \Cref{fig:qutrit-expsetup}, one of the outputs of the first TBS (after the circulator) is directly connected to the single-photon detector $\mathrm{D}_0$, corresponding to $\ket{0}$ in the path basis. The other output is connected to the second TBS, whose outputs are then connected to single-photon detectors $\mathrm{D_1}$ and $\mathrm{D_2}$ respectively, corresponding to $\ket{1}$ and $\ket{2}$. The phase modulators are controlled by a function generator which is connected to a Field Programmable Gate Array (FPGA) based pulse generator, which allows for dynamic control of the phase modulators during operation. We note that in order to remove the coherence between the Fock states that constitute the weak coherent states, the phase of each launched pulse should be randomized \cite{nie_experimental_2016}. However, for simplicity, we omit this in our proof-of-concept demonstration. Phase randomization is trivially implemented with, for example, a $\text{LiNbO}_3$ phase modulator in the source device.

The advantage of the measurement on the computational basis states lies in its simplicity, as it is independent of the relative phases $\varphi_1$ and $\varphi_2$ in \Cref{eq:qutrit-full-state}, thus making the experiment very stable. Furthermore, this can be directly scaled to even higher dimensions by adding more tunable beamsplitters in this fashion.


\section{Measurement Device Independent Randomness Generation}

\label{appendix:methods-certification}
In this section we outline the general structure of an MDI-QRNG protocol and then later specify the concrete instances employed in this work. 

The protocol operates in a prepare-and-measure setting with two parties, Alice (trusted) and Bob (untrusted). In each round, Alice selects a classical input $x$ based on a prior probability distribution $p(x)$ and prepares a quantum state $\rho_x$ on a known, fixed Hilbert space $\mathcal H$. The state is sent through a lossy and noisy quantum channel to Bob, who applies an uncharacterized measurement and outputs a classical outcome $a$. This yields conditional statistics $p(a|x)$.

In the MDI scenario, the entire detection apparatus and Bob's measurement (including the channel) are treated as black boxes and thus are untrusted; only Alice's preparation device is trusted and characterized. Operationally, the protocol alternates between \emph{test} ($T$) rounds and \emph{generation} ($G$) rounds. In $T$ rounds, Alice samples $x$ from a finite test set $X_T$ ($\{1,\ldots,|X_T|\}$) and prepares one of the known test states $\{\rho_x \in B_+(\mathcal H)\}_{x\in X_T}$ to probe Bob's device, where $B_+(\mathcal{H})$ is the space of bounded positive semi-definite operators acting on the Hilbert space $\mathcal{H}$. In the $G$ rounds, Alice prepares a state $\rho_x\in B_+(\mathcal H)$ with $x=G$, where $G=|X_T|+1$, intended for randomness generation. 

We assume an all-powerful but passive eavesdropper, Eve, who designs and manufactures Bob’s entire detection block (including the channel optics) before the protocol starts. She may choose arbitrary measurement operators, loss behavior, double-click rules and other classical post-processing; embed hidden classical randomness and long-term memory; and even keep a purification/quantum side system of any ancillas used in the device. Once deployed, however, Eve does not inject additional signals or interact adaptively with the run beyond what her prebuilt device does to the incoming quantum states; 

Specifically, let the channel be described by a CPTP (completely positive, trace-preserving) map $\mathcal{E}:B(\mathcal{H})\to B(\mathcal{H}_B)$, where $\mathcal{H}_{B}$ is the Hilbert space on which Bob's device implements a joint-measurement described by a POVM $\{N_{a,e}\}_{a,e}$, where $e$ is Eve's guess, such that $N_{a,e}\succeq0$ for all $a,e$, and $\sum_{a,e}N_{a,e}=\mathbb{I}_{\mathcal{H}_{B}}$, such that the observed statistics follow $p(a|x)=\sum_e\tr[\mathcal{E}(\rho_x)N_{a,e}]$. Switching to the Heisenberg picture with the unital adjoint $\mathcal{E}^*$ of $\mathcal{E}$, we define the effective joint POVM $\{M_{a,e}\}_{a,e}$ on $\mathcal{H}$, such that $M_{a,e}:=\mathcal{E}^*(N_{a,e})\succeq 0$ for all $a,e$, and  $\sum_{a,e}M_{a,e}=\mathbb{I}_{\mathcal{H}}$, such that $p(a|x)=\sum_e\tr[\rho_xM_{a,e}]$.

Up to this point the treatment is completely general—Eve is unrestricted. All physically allowed eavesdropping strategies (arbitrary channels, instruments, and measurements, plus classical post-processing) are captured by suitable choices of the effective POVM $\{M_{a,e}\}_{a,e}$ on $\mathcal H$. (Any such $\{M_{a,e}\}$ is also physically realizable via Stinespring–Naimark dilation \cite{pellonpaa_naimark_2023}). To restrict Eve to \emph{passive} attacks, we impose the following ``no-signaling'' (state-independent marginal) constraint on her guess:
\begin{equation} \label{noSig}
    \sum_aM_{a,e}=q(e)\mathbb{I}_\mathcal{H},
\end{equation}
for values of Eve's guess $e$ such that $\sum_eq(e)=1$ and $q(e)\geq 0$. In particular, the condition \eqref{noSig} implies that Eve's guess $e$ is independent of the particular state Alice prepares, that is, $p(e|x)=q(e)$ for all $x$. Eve's success probability is given by, 
\begin{equation} \label{GP}
     p_{\mathrm{guess}}=\sum_a \tr[\rho_xM_{a,a}].
\end{equation}
Finally, we put everything together to arrive at the following semi-definite program which maximizes Eve's guessing probability \eqref{GP} over all possible passive eavesdropping strategies, 
\begin{equation}
\label{eq:optimization-program}
\begin{aligned}
p^*_{\text{guess}}:=\max_{M,q}\;& \sum_{a}\,\Tr[\rho_x M_{a,a}]\\
\text{s.t. }\;& M_{a,e}\succeq 0\ (\forall a,e),\quad \sum_a M_{a,e}=q(e)\,\mathbb{I}_{\mathcal{H}}\ (\forall e),\\
& q(e)\ge 0\ (\forall e),\ \sum_{e} q(e)=1,\\
& \sum_{e}\Tr[\rho_x M_{a,e}]=p(a|x)\ (\forall x,e).
\end{aligned}
\end{equation}
The certified randomness as measured by per-generation-round min-entropy is given by
\begin{equation} \label{MinE}
    h=-\log_2(p^*_{guess}).
\end{equation}
We have described a general MDI--QRNG protocol where we modeled the experimental data by conditional probabilities $p(a|x)$. This is an idealization valid only in the asymptotic limit of infinitely many rounds. Next, we describe how to correct for finite rounds.

\section{Finite-round correction}
\label{appendix:methods-finite-round-correction}
In the finite-round regime we observe empirical frequencies $\hat p(a|x)$ that fluctuate around the true $p(a|x)$. For input $x$, let $n_x$ be the number of rounds in which $x$ was used, and let $\hat p(a|x)\;=\;\frac{N(a,x)}{n_x}$. where $N(a,x)$ is the number of occurrences of outcome $a$ given input $x$. By the (two-sided) Chernoff--Hoeffding inequality \cite{supic_measurement-device-independent_2017},
\begin{equation}
p\!\Big(\,\big|\hat p(a|x)-p(a|x)\big|\ge t_x(\varepsilon_{x,a})\,\Big)\;\le\;\varepsilon_{x,a},
\end{equation}
with
\begin{equation}
t_x(\varepsilon_{x,a})\;=\;\sqrt{\frac{\ln\!\big(2/\varepsilon_{x,a}\big)}{2\,n_x}}.
\end{equation}
Equivalently, with probability $\varepsilon_{x,a}$,
\begin{equation}
\label{eq:CH-interval}
\hat p(a|x)-t_x(\varepsilon_{x,a})\;\le\;p(a|x)\;\le\;\hat p(a|x)+t_x(\varepsilon_{x,a}).
\end{equation}

We incorporate~\eqref{eq:CH-interval} into the SDP by replacing the equality constraints with linear inequalities:
\begin{equation}
L_{x,a}\;\le\;\sum_{e\in A}\tr\!\big[\rho_x M_{a,e}\big]\;\le\;U_{x,a},
\end{equation}
for all $x,a$, where $L_{x,a}:=\max\{0,\hat p(a|x)-t_x\}$ and $U_{x,a}:=\min\{1,\hat p(a|x)+t_x\}$.

In the MDI setting, an eavesdropper can exploit detection behavior—losses (no clicks) and multi-click events—to make the protocol appear secure while still \emph{perfectly predicting} Bob’s outputs. Simply discarding rounds with no clicks or multiple clicks amounts to a fair-sampling assumption and leaves open the so-called \emph{detection loophole}, which can be leveraged to bias the retained data and fake security. Next, we discuss the detection loophole in the context of MDI–QRNG and describe how our methodology prevents it.

\section{Detection loophole}
\label{appendix:methods-detection-loophole}
Losses and multi–click events are inevitable in experiments involving quantum optics. In an MDI–QRNG, where the measurement device is untrusted, an adversary can exploit detection-behavior (by tailoring double-click and no-click rules) to make the protocol appear secure while rendering the outcomes predictable. Prior works (e.g., \cite{carine_multi-core_2020,argillander_all-fiber_2022,alarcon_dynamic_2023,Argillander2023}) discarded no-click and multi-click events, which amounts to a fair-sampling assumption and leaves a detection loophole.

The most general way to avoid this loophole is to treat \emph{every} physically distinct detection pattern as a separate measurement outcome. For an idealized device with $D$ binary detectors, let the outcome alphabet be $A \;=\; \{0,1\}^{D}$,
so that $a=(a_1,\dots,a_D)\in A$ encodes the click pattern ($a_i=1$ means detector $i$ clicked, $a_i=0$ means it did not). The no-click event is $a=\mathbf{0}$, single-clicks have Hamming weight $\|a\|_1=1$, and multi-clicks have $\|a\|_1\ge 2$. We then collect and use the full conditional statistics $p(a|x)$ with $a\in A$ \emph{without post-selection}. In the semi-definite program \eqref{eq:optimization-program}, this simply enlarges the outcome set: the effective joint POVM becomes $\{M_{a,e}\}_{a,e\in A}$. This treatment certifies randomness while explicitly accounting for all adversarial strategies that exploit losses or manipulate click patterns.

We have laid the general framework MDI-QRNG protocols. A particular MDI-QRNG protocol is instantiated by specifying: $(i.)$ trusted preparation device: the set of fully characterized Alice's states $\{\rho_x\in B_+(\mathcal{H}))\}$, including the Hilbert space $\mathcal{H}$; and $(ii.)$ the measurement outcome alphabet $A$ determined by the number of detectors $D$ involved in the experiment. Next, we present the specifications of the experimental MDI-QRNG protocols featured in this work. 

\section{Experimental Qutrit MDI-QRNG}
\label{appendix:methods-experimental-qutrit-mdi-qrng}
To model Alice’s path-encoded weak coherent states (WCS) produced by heavily attenuated laser pulses, we take the preparation Hilbert space to be the three-mode bosonic Fock space $\mathcal H \;=\; \mathcal F(\mathbb C^{3})$,
with Fock basis $\{\ket{n_1n_2n_3}\,:\,n_1,n_2,n_3\in\mathbb{N}_{0}\}$. 

For each setting $x$ (three test states $x\in X_T=\{1,2,3\}$ and one generation state $x=4$),
let the amplitudes be $\boldsymbol\alpha_x=(\alpha_{1|x},\alpha_{2|x},\alpha_{3|x})$ with total mean photon number
$\mu:=\sum_{i=1}^3|\alpha_{i|x}|^2$ and normalized mode vector
$\boldsymbol\beta_x:=\boldsymbol\alpha_x/\sqrt{\mu}$.
With global phase randomization the state is block-diagonal in total photon number:
\begin{equation}
\rho_x \;=\; \sum_{n=0}^{\infty} p(n)\,\ket{\psi_x^{(n)}}\!\bra{\psi_x^{(n)}},
\label{eq:wcs-block}
\end{equation}
where $p(n)=e^{-\mu}\frac{\mu^n}{n!}$ and
\begin{equation}
\ket{\psi_x^{(n)}} \;=\; \frac{1}{\sqrt{n!}}\Big(a^\dagger(\boldsymbol\beta_x)\Big)^n\ket{000},
\label{eq:wcs-n}
\end{equation}
where 
\begin{equation}
    a^\dagger(\boldsymbol\beta_x):=\sum_{i=1}^3 \beta_{i|x}\,a_i^\dagger.
\end{equation}
Since the complexity of the SDP in~\eqref{eq:optimization-program} scales with $\dim\mathcal H$, we truncate to the three-photon sector 
\begin{equation}
    \mathcal H' \;:=\; \mathrm{span}\{\ket{n_1n_2n_3}:n_1,n_2,n_3 \in \mathbb{N}_0 \ \& \ n_1+n_2+n_3\le 3\},
\end{equation}
which has dimension 
\begin{equation}
    \dim\mathcal H'=\binom{3+3}{3}=20.
\end{equation}
Let $\Pi\equiv\Pi_{\le 3}$ denote the projector onto $\mathcal H'$. We now define the in-subspace weight and decomposition as
\begin{equation}
    \kappa:=\Tr[\Pi\rho_x]=e^{-\mu} \sum^3_{n=0}\frac{\mu^n}{n!},
\end{equation}
and $\rho_x=\kappa\,\rho_x' + (1-\kappa)\,\tau_x^{\perp}$, where $\rho_x'=\Pi\rho_x\Pi/\kappa_x$ supported on $\mathcal H'$ and
$\mathrm{supp}(\tau_x^{\perp})\subseteq\mathcal H'^{\perp}$. 

We run the SDP on $\mathcal H'$ using the renormalized states $\rho_x'$. Let $p^{*'}_{\text{guess}}$ be Eve's optimal guessing probability returned from \eqref{eq:optimization-program}. The worst-case \emph{physical} guessing probability then satisfies 
\begin{equation} \label{TruncCorrection}
    p^{*}_{\text{guess}}=\kappa p^{*'}_{\text{guess}}+(1-\kappa),
\end{equation}
which is used to compute randomness via \eqref{MinE}. 

We choose three path-orthogonal test settings with all photons in a single mode:
\begin{align}
   \ket{\psi^{(n)}_1}&=\ket{n00} \\
   \ket{\psi^{(n)}_2}&=\ket{0n0} \\
   \ket{\psi^{(n)}_3}&=\ket{00n},
\end{align}
for $n\in\{0,1,2,3\}$.
For the generation state we take the equal superposition
$\boldsymbol\beta_4=(1,1,1)/\sqrt{3}$, so
$a^\dagger(\boldsymbol\beta_4)=\tfrac{1}{\sqrt{3}}(a_1^\dagger+a_2^\dagger+a_3^\dagger)$ and
\begin{align}
    \ket{\psi^{(0)}_4}&=\ket{000},\\
    \ket{\psi^{(1)}_4}&=\frac{1}{\sqrt{3}}\big(\ket{100}+\ket{010}+\ket{001}\big)\\
    \ket{\psi^{(2)}_4}&=\frac{1}{3}\big(\ket{200}+\ket{020}+\ket{002}\big) \\
    & \ \ \ + \frac{2}{3\sqrt{2}}\big(\ket{110}+\ket{101}+\ket{011}\big)\\
    \ket{\psi^{(3)}_4}&=\frac{\sqrt{3}}{9}\big(\ket{300}+\ket{030}+\ket{003}\big)\\
    & \ \ \ + \frac{1}{3}\big(\ket{210}+\ket{120}+\ket{201} + \ket{102}\\
    & \quad \quad \quad \ \ +\ket{021}+\ket{012}\big) + \frac{2}{3\sqrt{2}}\ket{111}
\end{align}

Finally, in our experiment we employ $D=3$ binary detectors, so the outcome alphabet is $A=\{0,1\}^3$. An outcome $a=(a_1,a_2,a_3)\in A$ encodes the click pattern. We model Bob's untrusted device by an effective joint POVM $\{M_{a,e}\}_{a,e\in A}$ on $\mathcal H$, where $e\in A$ is Eve’s guess.

Next, we describe the qubit MDI-QRNG protocol obtained via post-processing the outcomes of the qutrit experiment.

\section{(Binned) Qubit MDI-QRNG}
\label{appendix:methods-binning}
To obtain the qubit MDI-QRNG from the experimental qutrit MDI-QRNG described above, we simply ignore one of the three paths and the associated detector.

We consider a qubit MDI--QRNG protocol with two test states 
($x'\in X'_T=\{1,2\}$) and one generation state ($x'=3$). 
The preparation Hilbert space is the two-mode bosonic Fock space 
$\mathcal H=\mathcal F(\mathbb C^2)$ with basis $\{\ket{n_1n_2}:n_1,n_2\in\mathbb N\}$.

As in the qutrit case, global phase randomization makes the states block-diagonal 
in the total photon number:
\begin{equation}
\rho_{x'} \;=\; \sum_{n=0}^\infty p(n)\,\ket{\psi_{x'}^{(n)}}\!\bra{\psi_{x'}^{(n)}},
\label{eq:wcs-block2}
\end{equation}
where $p(n)=e^{-\mu}\frac{\mu^n}{n!}$, $\mu=|\alpha_{1|x'}|^2+|\alpha_{2|x'}|^2$ is the mean photon number, and
\begin{equation}
\ket{\psi_{x'}^{(n)}} \;=\; \frac{1}{\sqrt{n!}}
\Big(a^\dagger(\boldsymbol\beta_{x'})\Big)^n\ket{00},
\label{eq:wcs-n2}
\end{equation}
where $a^\dagger(\boldsymbol\beta_{x'})=\beta_{1|{x'}}a_1^\dagger+\beta_{2|{x'}}a_2^\dagger$ with normalized mode vector $\boldsymbol\beta_{x'}=\boldsymbol\alpha_{x'}/\sqrt{\mu}$. 

Analogously to the qutrit case, we truncate to the three-photon sector 
\begin{equation}
    \mathcal H' \;:=\; \mathrm{span}\{\ket{n_1n_2}:\ n_1,n_2\in \mathbb{N}_0 \ \& \ n_1+n_2\le 3\},
\end{equation}
which has dimension 
\begin{equation}
    \dim\mathcal H'=\binom{3+2}{2}=10,
\end{equation}
and we similarly correct the guessing probability for the truncation. 

We choose two path-orthogonal test set settings with all photons in a single mode:
\begin{align}
   \ket{\psi^{(n)}_1}&=\ket{n0} \\
   \ket{\psi^{(n)}_2}&=\ket{0n},
\end{align}
for $n\in\{0,1,2,3\}$. For the generation state we take the equal superposition $\boldsymbol\beta_4=(1,1)/\sqrt{2}$, so
$a^\dagger(\boldsymbol\beta_3)=\tfrac{1}{\sqrt{2}}(a_1^\dagger+a_2^\dagger)$ and
\begin{align}
    \ket{\psi^{(0)}_3}&=\ket{00},\\
    \ket{\psi^{(1)}_3}&=\frac{1}{\sqrt{2}}\big(\ket{10}+\ket{01}\big)\\
    \ket{\psi^{(2)}_3}&=\frac{1}{2}\big(\ket{20}+\ket{02}\big) +\frac{1}{\sqrt{2}}\ket{11} \\
    \ket{\psi^{(3)}_3}&=\frac{1}{2\sqrt{2}}\big(\ket{30}+\ket{03}\big)+ \frac{\sqrt{6}}{4 }\big(\ket{21}+\ket{12}\big)
\end{align}

Finally, since we keep only two of the three detectors, the outcome alphabet becomes $A'=\{0,1\}^2$. We implement a fixed classical coarse–graining termed ``binning" that marginalizes over the third detector. 
Writing $a=(a_1,a_2,a_3)\in\{0,1\}^3$ and
$a'=(a_1',a_2')\in A'$, define
\begin{equation}
    T(a'|a):=\mathbf 1\{(a_1',a_2')=(a_1,a_2)\}.
\end{equation}
Then the observed data transforms linearly as,
\begin{align}
    p'(a'|x')&=\sum_{a\in\{0,1\}^3} T(a'|a)\,p(a|x=x')\\
&=\sum_{a_3\in\{0,1\}} p(a_1',a_2',a_3\,|\,x=x'),
\end{align}
The no–click event on the retained detectors remains explicit as $a'=(0,0)$,
so the treatment remains detection-loophole free for the kept detection block. 

We note here that the observed data $p'(a'|x')$ thus obtained data only remains consistent with mean photon number for the qutrit case $\mu'=\mu$ for the test states and with a scaled-down mean photon number $\mu'=\frac{2}{3}\mu$ for the generation state. To consistently compare with the qutrit MDI-QRNG, we require the mean photon number for all states, so we apply a fixed, state-independent loss map to the test-state data with retention
probability $r=\tfrac{2}{3}$. 

Defining $A'=\{0,1\}^2$ and
\begin{equation}
    T_{\mathrm{loss}}(b|a')=
\begin{cases}
1, & a'=(0,0),\ b=(0,0),\\
r, & a'\neq (0,0),\ b=a',\\
1-r, & a'\neq (0,0),\ b=(0,0),\\
0, & \text{otherwise},
\end{cases}
\end{equation}
and we set,
\begin{equation}
    \tilde p(a'|x')=\sum_{c\in A'}T_{\mathrm{loss}}(a'|c)\,p'(c|x'),
\end{equation}
for all $x'\in X'_T$.

This preserves normalization while enforcing
consistency with $\mu'=\tfrac{2}{3}\mu$ on test settings. Since the operation is purely classical post-selection free post-processing, it keeps the analysis detection-loophole free and yields a conservative security bound.

Next, we present theoretical simulations for the qubit and qutrit MDI--QRNG implementations described above entailing specifically theoretical models for the measurement apparatus.

\section{Simulations}
\label{appendix:methods-simulations}
In this section, we specify the theoretical model used to accurately reproduce the experimental behaviors of the qutrit and qubit MDI--QRNGs. Concretely, we describe here the theoretical model to generate the simulated statistics $\tilde{p}(a|x)$. We then feed the simulated statistics into \eqref{eq:optimization-program} while keeping everything else same as described above.

We aim to capture the dominant experimental imperfections—specifically, background noise, dark-counts and full click patterns (including losses and multi-clicks). Background noise is modeled via a visibility parameter $\nu\in[0,1]$, yielding effective preparations
\begin{equation}
\tilde \rho_x \;=\; \nu\,\rho_x \;+\; (1-\nu)\,\frac{\mathbb I_{\mathcal H}}{\dim\mathcal H},
\end{equation}
where $\mathcal{H}$ represents the three-mode Fock space truncated to the three photon sector such that $\dim\mathcal H=20$ for qutrits, and for the qubit case $\mathcal{H}$ is the two-mode Fock space truncated to the three photon sector such that $\dim\mathcal H=10$. 

Next, we describe the measurement operators used to reproduce the full click pattern, including multiple clicks and no clicks.

Let us first consider the qutrit case. We model a device with three threshold detectors of efficiencies $(\eta_1,\eta_2,\eta_3)$ and per-window dark-count probabilities $(d_1,d_2,d_3)$. In the ideal limit ($\eta_i=1$, $d_i=0$), the device implements the computational-basis qutrit projective measurement with three outcomes. In the realistic model, the outcome alphabet is $A=\{0,1\}^3$, where $a=(a_1,a_2,a_3)\in A$ encodes the click pattern ($a_i=1$ iff detector $i$ clicks). We consider a POVM $\{M_{a_1,a_2,a_3}\}\subset B_{+}(\mathcal H)$ acting on the three-mode Fock space truncated to the $20$ dimensional $\le 3$-photon sector, 
$\mathcal H=\mathrm{span}\!\left\{\ket{n_1 n_2 n_3}:\ n_1,n_2,n_3\in\mathbb N_0,\ \& \ n_1+n_2+n_3\le 3\right\}$.

Assuming independence across modes and no crosstalk, the POVM elements are diagonal in the Fock basis and given by
\begin{align}
\label{eq:qutrit-povm-dark}
M_{a}
&=\!\!\sum_{\substack{n_1,n_2,n_3\ge 0\\ n_1+n_2+n_3\le 3}}
\Bigg(\prod_{i=1}^{3} p_i(a_i\,|\,n_i)\Bigg)\;
\ket{n_1 n_2 n_3}\!\bra{n_1 n_2 n_3}
\end{align}
where,
$p_i(1\,|\,n_i)=1-(1-d_i)(1-\eta_i)^{n_i}$,
$p_i(0\,|\,n_i)=(1-d_i)(1-\eta_i)^{n_i}$ for $i\in\{1,2,3\}$.
By construction $M_{a_1,a_2,a_3}\succeq 0$ and 
$
    \sum_{a\in\{0,1\}^3} M_{a}=\mathbb I_{\mathcal H}$.

Consequently, the simulated statistics follow from the Born rule,
\begin{equation}
    \tilde{p}(a|x)=\tr[\tilde{\rho}_xM_{a}],
\end{equation}
for all $x\in\{1,2,3,4\}$

Finally, we describe the analogous modelling of measurement operators for the qubit MDI--QRNG. We model a device with two threshold detectors of efficiencies $(\eta_1,\eta_2)$ and per-window dark-count probabilities $(d_1,d_2)$. In the ideal limit ($\eta_i=1$, $d_i=0$) the device implements the computational-basis qubit projective measurement with two outcomes. In the realistic model, the outcome alphabet is $A=\{0,1\}^2$, where $a=(a_1,a_2)\in A$ encodes the click pattern ($a_i=1$ iff detector $i$ clicks). We consider a POVM $\{M_{a_1,a_2}\}\subset B_{+}(\mathcal H)$ acting on the two-mode Fock space truncated to the $10$ dimensional $\le 3$-photon sector, $\mathcal H=\mathrm{span}\!\left\{\ket{n_1 n_2}:\ n_1,n_2\in\mathbb N_0,\ \& \ n_1+n_2\le 3\right\}$.

Assuming independence across modes and no crosstalk, the POVM elements are diagonal in the Fock basis and given by
\begin{align}
\label{eq:qubit-povm-dark}
M_{a}
&=\!\!\sum_{\substack{n_1,n_2\ge 0\\ n_1+n_2\le 3}}
\Bigg(\prod_{i=1}^{2} p_i(a_i\,|\,n_i)\Bigg)\;
\ket{n_1 n_2}\!\bra{n_1 n_2},\\
p_i(1\,|\,n_i)
&=1-(1-d_i)(1-\eta_i)^{n_i},\qquad\\
p_i(0\,|\,n_i)&=(1-d_i)(1-\eta_i)^{n_i}\quad(i=1,2).
\end{align}
By construction $M_{a}\succeq 0$ and $\sum_{a\in\{0,1\}^2} M_{a}=\mathbb I_{\mathcal H}$.

The simulated statistics follow from the Born rule,
\begin{equation}
    \tilde{p}(a|x)=\tr[\tilde{\rho}_x M_{a}],
\end{equation}
for all $x\in\{1,2,3\}$.

We feed these statistics back into the \eqref{eq:optimization-program} to obtain Eve's guessing probability $p^*_{\text{guess}}$ and min-entropy via \eqref{MinE}.

\begin{acknowledgments}
We acknowledge helpful discussions with Alvaro Alarc\'{o}n. This work was supported by the Wallenberg Center for Quantum Technologies. This work was partially supported by the Foundation for Polish Science (IRAP project, ICTQT, contract No. MAB/218/5, co-financed by EU within the Smart Growth Operational Programme). M.P., and P.R.D. acknowledge support from the NCN Poland, ChistEra-2023/05/Y/ST2/00005 under the project Modern Device Independent Cryptography (MoDIC). A.C. acknowledges financial support by NCN grant SONATINA 6 (contract No. UMO-2022/44/C/ST2/00081). This work is partially carried out under IRA Programme, project no. FENG.02.01-IP.05-0006/23, financed by the FENG program 2021-2027, Priority FENG.02, Measure FENG.02.01., with the support of the FNP.
\end{acknowledgments}

\section*{Author Contributions}
J.A., P.R.D., A.C. and G.B.X. conceived the idea. J.A., D.S.-L., and M.C. built the experimental setup and performed the measurements.  A.C. and P.R.D. developed the theoretical model and simulations with M.P.. J.A., A.C. and G.B.X. analysed the data. G.B.X. supervised the project. All authors contributed to writing the manuscript.

\bibliography{sample.bib}

\end{document}